# Structural and magnetic properties of molecular beam epitaxy $(MnSb_2Te_4)_x(Sb_2Te_3)_{1-x}$ topological materials with exceedingly high Curie temperature

Candice R. Forrester,[1,2,3] Christophe Testelin,[4] Kaushini Wickramasinghe,[1] Ido Levy,[5] Dominique Demaille,[4] David Hrabovski,[6] Xiaxin Ding,[7] Lia Krusin-Elbaum,[7,8] Gustavo E. Lopez,[2,3] Maria C. Tamargo[1,2]

[1] *Department of Chemistry, The City College of New York, NY, NY 10031*

[2] *PhD Program in Chemistry, CUNY Graduate Center, NY, NY 10016*

[3] *Department of Chemistry, Lehman College, Bronx, NY 10468*

[4] *Sorbonne Université, CNRS, Institut des NanoScience de Paris, F-75005 Paris, France*

[5] *Department of Physics, New York University, NY, NY 10003*

[6] *Sorbonne Université, MPBT Platform, 4 Place Jussieu, 75252 Paris, France*

[7] *Department of Physics, The City College of New York, NY, NY 10031*

[8] *PhD Program in Physics, CUNY Graduate Center, NY, NY 10016*

**Abstract**

Tuning magnetic properties of magnetic topological materials is of interest to realize elusive physical phenomena such as quantum anomalous hall effect (QAHE) at higher temperatures and design topological spintronic devices. However, current topological materials exhibit Curie temperature ($T_C$) values far below room temperature. In recent years, significant progress has been made to control and optimize $T_C$, particularly through defect engineering of these structures. Most recently we showed evidence of $T_C$ values up to 80K for $(MnSb_2Te_4)_x(Sb_2Te_3)_{1-x}$, where $0.7 \leq x \leq 0.85$, by controlling the compositions and Mn content in these structures. Here we show further enhancement of the $T_C$, as high as 100K, by maintaining high Mn content and reducing the growth rate from 0.9 nm/min to 0.5 nm/min. Derivative curves reveal the presence of two $T_C$ components contributing to the overall value and propose $T_{C1}$ and $T_{C2}$ have distinct origins: excess Mn in SLs and Mn in $Sb_{2-y}Mn_yTe_3$ QLs alloys, respectively. In pursuit of elucidating the mechanisms promoting higher Curie temperature values in this system, we show evidence of structural disorder where Mn is occupying not only Sb sites but also Te sites, providing evidence of significant excess Mn and a new crystal structure: $(Mn_{1+y}Sb_{2-y}Te_4)_x(Sb_{2-y}Mn_yTe_3)_{1-x}$. Our work shows progress in understanding how to control magnetic defects to enhance desired magnetic properties and the mechanism promoting these high $T_C$ in magnetic topological materials such as $(Mn_{1+y}Sb_{2-y}Te_4)_x(Sb_{2-y}Mn_yTe_3)_{1-x}$.

# INTRODUCTION

Three dimensional (3D) topological insulators (TIs) continue to gain attention for their unique properties, which are advantageous for novel electronic applications, including quantum computing[3,4] and spintronics.[1,2] Earlier work demonstrated quantum anomalous Hall effect (QAHE) in 2D systems, where electrons can exhibit dissipationless transport in the absence of a magnetic field,[5] led to the discovery of these materials, which employ spin-orbit coupling to affect electron transport in the crystal. The so-called 3D TIs, with molecular formulas $V_2VI_3$, in which the group V elements are Bi or Sb, and the group VI elements are Se and Te, emerged as a new class of materials with inverted conduction and valence bands as a result of the strong spin-orbit coupling.[6] This was first explored theoretically by Kane and Mele,[6] who provided evidence of the $Z_2$ topological invariant that distinguished ordinary insulators from topological insulators. Due to the strong spin-orbit coupling found in these materials, they are highly efficient in spin to charge conversions[1,2] and do not suffer from back scattering, making them ideal materials for dissipationless transport.[7] Due to the topological nature of these materials, qubits made from TI's are expected to be robust to external perturbations and decoherence effects, and hence fault tolerant.[8] TIs are unique because they are comprised of insulating (semiconducting) bulk interiors and conducting Dirac surface states in which the electrons are massless and move along the surface in a spin-locked helical direction and are protected by time reversal symmetry (TRS).[9,10] Structurally, 3D TIs have a rhombohedral, R3m space group and self-assemble into quintuple layer (QL) units, held together via weak van der Waals forces.[11]

Unlike 3D TIs, magnetic topological insulators (MTIs) such as $MnSb_2Te_4$ experience broken TRS due to spontaneous magnetization introduced by magnetic ions such as Mn. The interaction between the surface states and the magnetic ions causes a gap to form at the Dirac point.[12,13] A gap at the Dirac point enables the observation of elusive physical phenomena like QAHE in these systems.[12,13,14] The incorporation of enough Mn, transforms the QL into a septuple layer (SL), with an added Mn layer in the center of the unit cell and another Te layer. Bulk $MnSb_2Te_4$ is intrinsically antiferromagnetic, which inhibits the realization of QAHE. However, a single SL is ferromagnetic (FM) with spins aligned out of plane.[15] Several approaches have been shown to induce FM coupling between SLs, including separating individual FM SLs with non-magnetic QLs,[16,17] growing an odd number of SLs[18] or introducing excess Mn, in the form of Mn antisites, into the all-SL structure[19, 20, 21].

For MnBi$_2$Te$_4$, separating FM SLs with non-magnetic QLs not only demonstrated QAHE but also resulted in a material with a Curie temperature (T$_C$) of about 15K.[17] In the case of excess Mn in MnSb$_2$Te$_4$, Wimmer[20] et al observed a T$_C$ of 50K, and most recently our group reported T$_C$ values of 80K[22] in structures of MnSb$_2$Te$_4$ SLs and Sb$_2$Te$_3$ QLs, which are the highest reported T$_C$ values for these materials systems. Structurally, Mn ions incorporate into a QL TI host crystal either as a dopant, substituting for one of the constituent atoms, or as a structural element forming the new SL crystal structure.[14, 21, 23] Both types of substitutions have significant effects on the properties of the final structure, and the mechanisms promoting the observed high T$_C$ are not well known. Many have proposed the presence of the Mn antisites in the SL as the main cause of the high T$_C$[19, 20, 21], while others have shown that alloy formation with transition metal ions (TMs) such as Sb$_{2-y}$TM$_y$Te$_3$ can also yield very high T$_C$.[24,25,26] Learning to tune these defects for higher T$_C$ values in this materials system can bring us closer to the design of practical spintronic devices using MTIs.

In this work we report on (MnSb$_2$Te$_4$)$_x$(Sb$_2$Te$_3$)$_{1-x}$ structures with exceptionally high T$_C$ values of more than 100K by further tuning the growth parameters, in particular by reducing the growth rate. These values are by far the highest T$_C$ observed for this materials system. Temperature dependent Hall resistance (R$_{xy}$) and magnetization measurements both reveal the high T$_C$. Other characterization techniques such as X-ray diffraction rocking curves and energy-dispersive X-ray spectroscopy (EDS) were used to investigate the structural properties. We observed that samples grown at slow growth rates have more disorder, in the form of excess Mn, possibly occupying both Sb and Te sites. We propose that, for high Mn content samples, the Mn is incorporated in both in the SLs and the QLs and propose the formation of a new composite material: (Mn$_{1+y}$Sb$_{2-y}$Te$_4$)$_x$(Sb$_{2-y}$Mn$_y$Te$_3$)$_{1-x}$ that gives rise to the exceedingly high T$_C$ values.

**EXPERIMENTAL DETAILS**

All (MnSb$_2$Te$_4$)$_x$(Sb$_2$Te$_3$)$_{1-x}$ structures were grown by molecular beam epitaxy (MBE) on (0001) c-plane sapphire substrates in a Riber 2300P system under ultra-high vacuum conditions (3-5x10$^{-10}$ Torr). The MBE is equipped with reflection high-energy electron diffraction (RHEED) to monitor surface quality of the samples as they grow.

Before growth of $(MnSb_2Te_4)_x(Sb_2Te_3)_{1-x}$, the epi-ready sapphire substrates are baked at 670ºC for 1 hour to remove any impurities on the sapphire surface. Fluxes for 6 N antimony (Sb) with a Riber double zone cell, and single zone Knudsen cells for 6N tellurium (Te) and 5N8 manganese (Mn) were measured by the beam equivalent pressure (BEP) obtained by a gauge placed in the substrate position. The samples were grown with varying Mn BEP ratios, defined as $BEP_{Mn}/(BEP_{Mn} + BEP_{Sb})$, to control the $(MnSb_2Te_4)_x(Sb_2Te_3)_{1-x}$ composition (i.e, the value of x) and under excess Te, with a Sb:Te ratio of 1:20. Two growth rate ranges were used: fast growth rate of 0.9 – 1.0 nm/min and slow growth rate of ~ 0.4 - 0.5 nm/min. The two growth rates were obtained by adjusting the Sb flux. Most of the slow growth rate samples were grown with a Mn BEP ratio of 0.09, which led to the highest $T_C$ values for the fast growth rate samples.[22] For the growth, the substrate temperature ($T_{sub}$) was first raised to ~ 200ºC, where the Sb and Te shutters are opened to grow a low temperature buffer (LTB) layer. The $Sb_2Te_3$ LTB was grown for 8-10 minutes, until reconstruction of the RHEED was observed indicating crystalline $Sb_2Te_3$ layer formation. The $T_{sub}$ was then raised to 265ºC and the Mn shutter was opened to commence the growth of the desired layer of $(MnSb_2Te_4)_x(Sb_2Te_3)_{1-x}$. After 15 minutes of growth, the Sb and Mn shutters are closed and the $T_{sub}$ is raised to 300 ºC and held at this temperature for 15 minutes (15 minute anneal). After the 15 minute anneal, the Mn and Sb shutters are opened, and the remaining $(MnSb_2Te_4)_x(Sb_2Te_3)_{1-x}$ thin film is grown for 1 - 2 hours.[22]

After growth, samples were characterized by various techniques including atomic force microscopy (AFM) using a Bruker Dimension FastScan AFM with a FastScan-A silicon probe and high-resolution X-ray diffraction (HR-XRD) analysis using a Bruker D8 Discover diffractometer with a da Vinci configuration and a Cu Kα1(1.5418 Å) source. Further analysis of the (0012) X-ray rocking curves were measured using a crystal analyzer. Transport measurements were performed in the van der Pauw geometry with indium contacts using a 14 T Quantum Design physical property measurement system (PPMS) in a 1 mTorr (at low temperature) of He gas or in a Lakeshore 7600 electromagnetic system. Magnetization measurements of the samples were taken using a superconducting quantum interference device (SQUID, MPMS3 from Quantum Design). Energy-dispersive X-ray spectroscopy (EDS) measurements were made in a Zeiss Supra 40 with a compact 30mm Bruker detector.

**RESULTS and DISCUSSION**

Previous work by our group showed that the highest Curie temperature values for these materials, as high as 80K, were achieved in $(MnSb_2Te_4)_x(Sb_2Te_3)_{1-x}$, for values of x between 0.7 and 0.85 (70 – 85% SLs).[22] In that work, the highest $T_C$ values were obtained with a Mn BEP ratio of about 0.09 along with a growth temperature of 265 ºC and a 15 minute pre-annealing step. Typical growth rates for those materials were 0.9-1.0 nm/min, referred to in the remainder of this paper as 'fast growth rates'. Details of the growth can be found in the Supplementary Information of Ref. 27. In this work we will show that reducing the growth rate to ~0.4 - 0.5 nm/min ('slow growth rates') the Curie temperature has increased further to values above 100K. We first examine the structural and magnetic properties of these new slow growth rate materials to better understand the reason for the high $T_C$ values in these structures.

***Structural and electrical properties of $(MnSb_2Te_4)_x(Sb_2Te_3)_{1-x}$ with $x \geq 0.7$ grown with slow growth rate***

During the growth by molecular beam epitaxy, the surface was monitored *in-situ* using reflection high energy electron diffraction (RHEED) where the observed patterns indicate the crystalline and surface quality of the material as it grows. As in the case of the fast growth rate samples reported in Ref. 22, we observed streaky RHEED patterns for the slow growth rate samples, signifying good crystallinity and smooth surfaces. AFM images of the surfaces also indicate smooth surfaces, typical of the $Sb_2Te_3$ based materials, and their quality is unchanged with the changing growth rates. (See Supplementary Information, Figs. S1 and S2).

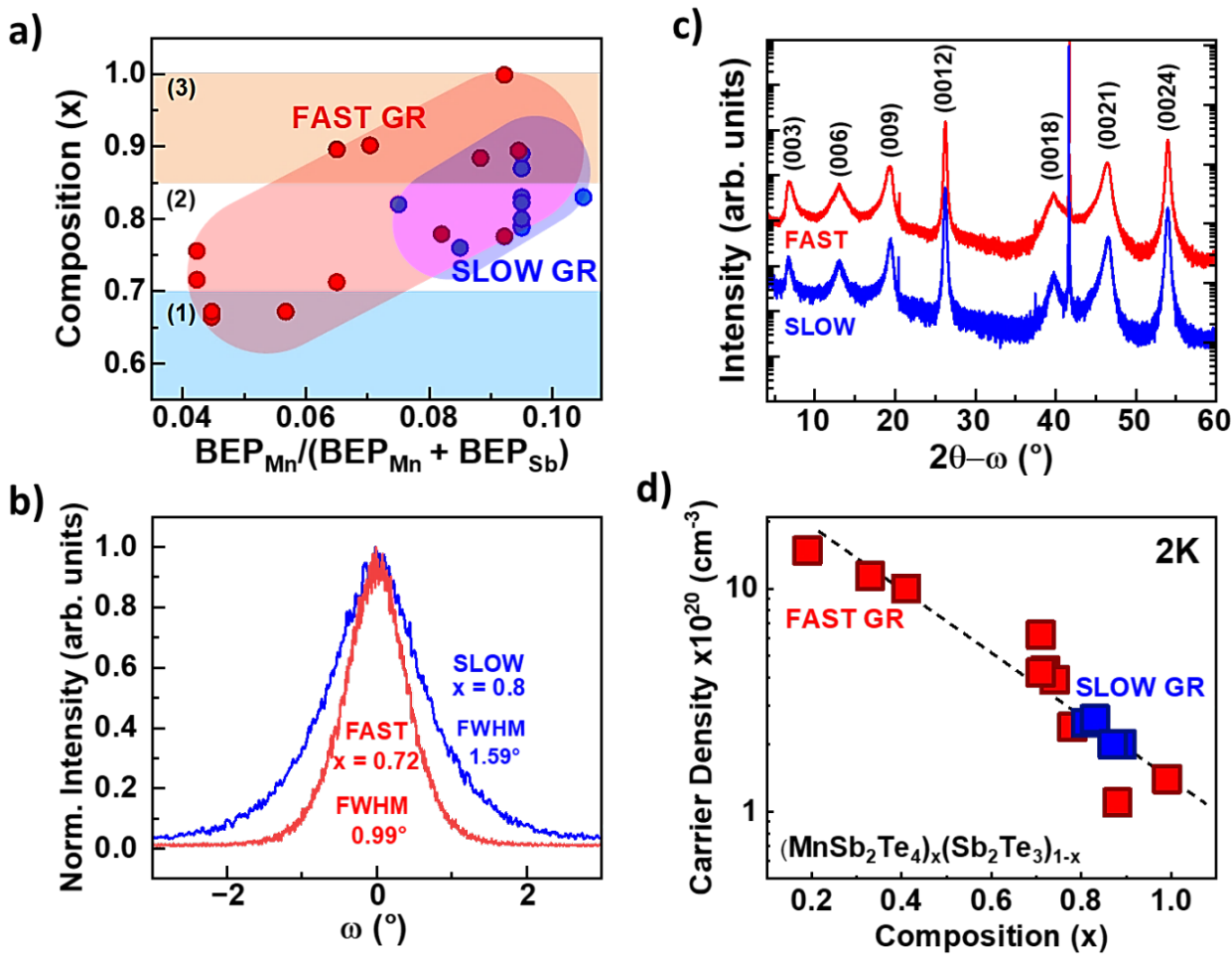

**Figure 1: Structural and electrical properties of $(MnSb_2Te_4)_x(Sb_2Te_3)_{1-x}$ structures grown at slow growth rates.** a) Fraction (x) of $MnSb_2Te_4$ as a function of Mn BEP ratio for samples grown with slow growth rates (blue points) superimposed on the corresponding data previously obtained for samples grown at a fast growth rate (red points). Based on the fraction (x), three composition regions are defined:[22] region 1, where x < 0.7 (light blue), region 2 where $0.7 \leq x \leq 0.85$ (white) and region 3 where x > 0.85 (orange). Blue and red shaded regions are used to highlight the relationship between the Mn BEP ratio and the fraction of $MnSb_2Te_4$. b) 2θ-ω X-ray diffraction scans for two samples of $(MnSb_2Te_4)_x(Sb_2Te_3)_{1-x}$, one grown at a slow growth rate (blue trace) and the other at a fast growth rate. c) Comparison of (0012) rocking curves of a slow growth rate (blue) and fast growth rate (red) samples. The broader (0012) peak measured for the slow growth rate sample indicates more disorder. d) Carrier density as a function of fraction (x) of $MnSb_2Te_4$ for fast (red squares) and slow (blue squares) growth rate samples. Higher fractions of $MnSb_2Te_4$ have lower carrier densities. Reducing the growth rate did not affect the carrier density for samples with the same SL fraction.
6

In this work, we aimed to grow samples at the slower growth rates with compositions $x \geq 0.7$ which are the compositions that previously yielded the higher $T_C$ values. The samples were characterized using high resolution X-ray diffraction (XRD) 2θ-ω scans to determine their composition x (or %SLs as described in Refs. 22 and 27). The composition was calculated based on the position of the (0015) peak of $Sb_2Te_3$ which evolves into the (0021) peak of $MnSb_2Te_4$.[27] Figure 1a summarizes the results of the slow growth rate samples grown, shown in blue circles, compared to previously reported compositions (red circles) for the fast growth rate samples.[22] As seen in Fig 1a, for the same Mn BEP ratios, slightly lower compositions are obtained for the samples grown at the slow growth rates; furthermore, we were unable to achieve 100% SL (or compositions of x = 1.0) using slow growth rates. Figure 1b compares the 2θ-ω scans of two samples having similar thickness of 53 nm and 66 nm, one grown at the fast growth rates (red) and another grown at the slow growth rates (blue), respectively. Based on the position of the (0021) peak in these samples, the calculated compositions are x = 0.72 and x = 0.80, respectively. To better assess the crystalline quality, a more sensitive study, using X-ray rocking curves, was performed to compare the quality of the fast and slow growth rate samples. Based on the full-width-half-maximum (FWHM) of the rocking curves shown in Fig. 1c (the slow growth rate sample has a FWHM of 1.59°, while for the fast growth rate sample it is 0.99°), we conclude that the slow growth rate produces material with similar composition but a higher degree of crystalline disorder. Finally, since Mn ions behave as electrical dopants in these materials, we also explored the impact of the slow growth rate on the electrical properties of these samples. Hall Effect measurements at high magnetic fields were performed for some of the samples grown at the slow growth rates and compared to previously reported values for samples grown at fast growth rates.[22] As seen in Figure 1d, the slow growth rate samples (blue squares) follow the same dotted line trend as the fast growth rate samples (red squares), suggesting that the slow growth rate does not have a significant effect on the samples' carrier density.

*Magnetic properties of $(MnSb_2Te_4)_x(Sb_2Te_3)_{1-x}$ with $x \geq 0.7$ grown at slower growth rate*

Field dependent Hall resistance was measured at 10 K for all the samples grown at slow growth rates. As was the case for the fast growth rate samples,[22] the plots all show typical hysteresis behavior signifying the ferromagnetic (FM) nature of all the samples investigated.

Ferromagnetic behavior was also observed in magnetization measurements made using a SQUID magnetometer. (See Supplementary Information, Fig. S3)

Figure 2 illustrates the temperature dependent Hall resistance and magnetization measurements of the slow growth rate samples for samples with composition 0.7 < x < 0.85. This composition range, which we previously called Region 2, [22] produced the highest $T_C$ values (as high as 80K) for fast growth rate samples. Figure 2a compares the normalized temperature dependent Hall resistance for two samples with similar compositions, one grown at slow growth rate and the other at the fast growth rate. The value at which the Hall resistance goes to zero is taken as the value of the $T_C$. According to Figure 2a, the fast growth rate sample has a $T_C$ of ~80K, as reported in Ref. 22, while the slow growth rate sample has value of ~100K, 25% higher than our previously reported value, and the highest $T_C$ value reported to date for these materials systems. This value is confirmed by magnetization measurement in a SQUID magnetometer where magnetic moment (M) was measured as a function of temperature also at 200G, shown in Figure 2b. A $T_C$ value of ~ 100K was observed from the magnetization measurement study, in agreement with the Hall resistance data. The inset in Fig. 2b shows a magnified view of the temperature dependent Hall resistance measurement of Fig 2a for the slow growth rate sample that shows the excellent agreement between the two measurements.

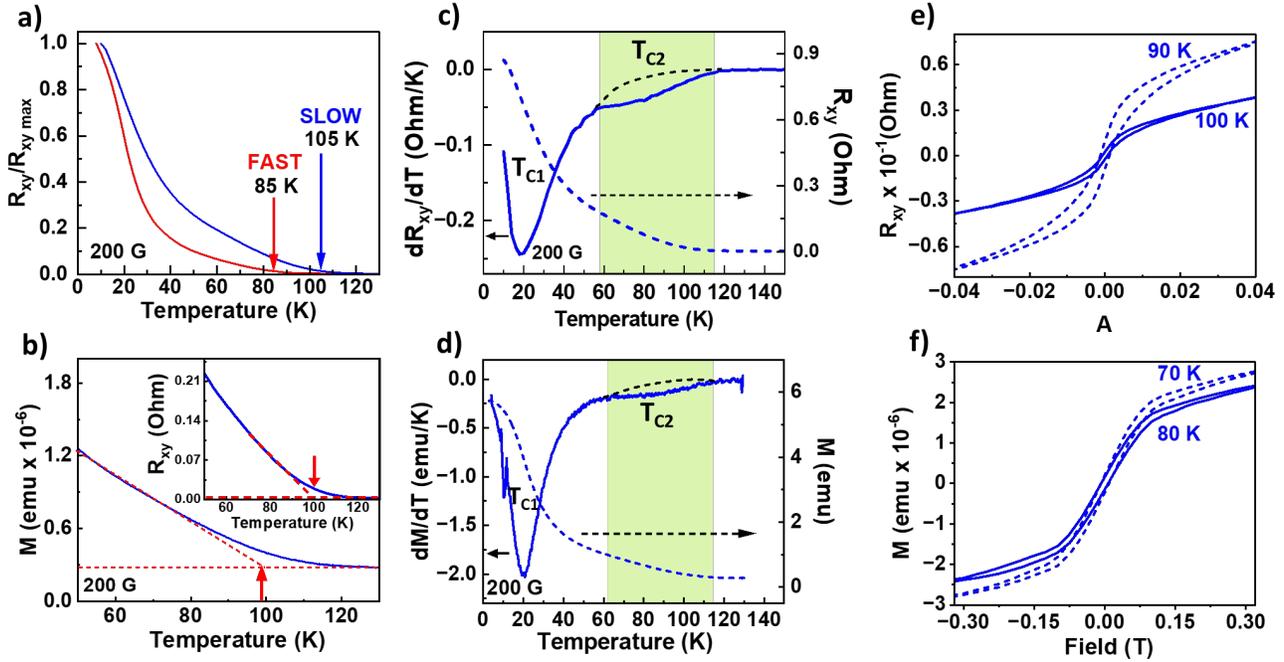

**Figure 2: Magnetic properties of $(MnSb_2Te_4)_x(Sb_2Te_3)_{1-x}$ structures with $0.7 \leq x \leq 0.85$ (Region 2).** a) Normalized temperature dependent Hall resistance ($R_{xy}$) at 200G of two samples, one grown at fast growth rates (red trace) and the other at slow growth rates (blue). The slow growth rate sample exhibited $T_C$ of about 100K. b) Temperature dependent magnetization measurements for the slow growth rate sample shown in Figure 2a. Good agreement is seen between the magnetization measurement and the Hall resistance measurement (inset). c) Superimposed plot of temperature dependent Hall resistance $R_{xy}$ (dashed) and its derivative curve (solid) for a sample grown at a slow growth rate. Two transitions are seen, indicating two $T_C$ components: one at ~ 20K and another reaching as high as 100K. d) Superimposed plot of temperature dependent magnetic moment (M) (dashed) and its derivative curve (solid) for the sample grown with a slow growth rate. Two $T_C$ components are seen, as in Fig. 2 c. e) Field dependent Hall resistance measurement for the slow growth rate sample with x = 0.83, showing hysteresis up to 100K. f) Field dependent magnetic moment (M) of the samples of Fig. 2e, showing hysteresis up to 80K.

To clearly establish the value of the $T_C$, derivative curves were plotted for both the Hall resistance ($R_{xy}$) and magnetization (M) data. In our earlier work, we showed that the derivative curves of the temperature dependent $R_{xy}$ gave evidence of two Curie temperatures, a low $T_C$ of about 20K and a high $T_C$ of 80K for the fast growth samples. Similarly, two distinct $T_C$ components could also be seen from the derivative plots of the slow growth rate samples shown here: a low $T_C$, $T_{C1}$, at ~20K and a second higher $T_C$, $T_{C2}$, with a broad, weaker signal reaching a value greater than 100K; this is shown in Figure 2c. Evidence for the two $T_C$ components is also obtained from the derivative plot of the magnetization data, shown in Figure 2d, with $T_C$ values in excellent agreement with the Hall resistance data. Finally, further evidence of the very high $T_{C2}$ value for the slow growth rate samples is obtained from the field dependent $R_{xy}$ plots, which show hysteresis loops that persist up to 100K, as shown in Figure 2e. Field dependent magnetization measurements were also made, as shown in Figure 2f. Hysteresis loops were evident up to about 80K (solid line) demonstrating the presence of a high $T_C$, although not as high as that measured from Rxy or as expected from the magnetization derivative plot. We believe that the suppression of the hysteresis loop at a lower temperature in the magnetization plots is probably due to the lower sensitivity of the magnetization probe.

We also investigated samples with composition x > 0.85, which we had previously identified as Region 3 in Ref. 22. At the fast growth rates[22], the samples in Region 3 had lower $T_C$ values than the samples in Region 2, with $T_C$ on the order of 40K. This is shown by the normalized temperature dependent Hall resistance data given by the red trace in Figure 3a. The figure also shows the result for a sample grown at slow growth rate (blue trace). The temperature dependent $R_{xy}$ plot suggests a high $T_C$ component ($T_{C2}$) of ~75K for the slow growth rate sample, significantly higher than the 40K observed for the fast growth rate sample. Derivative curves of the $R_{xy}$ plots of the slow growth rate sample (Figure 3b) show evidence of two $T_C$ components, and the field dependent $R_{xy}$ (Figure 3c) shows a $T_{C2}$ as high as 75K (solid line), suggesting a similar effect on the $T_C$ by the reduced growth rates in Region 3 samples.

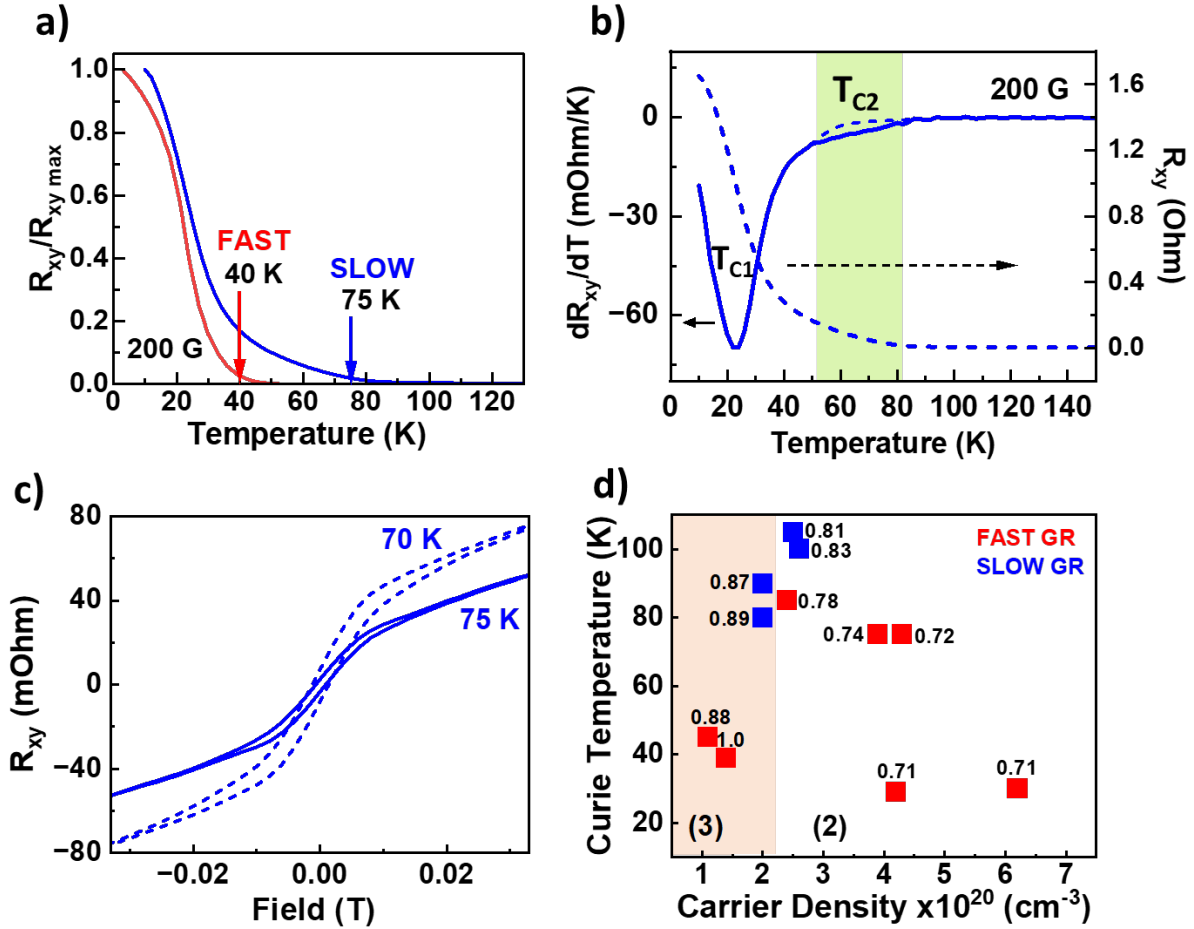

**Figure 3: Magnetic properties of $(MnSb_2Te_4)_x(Sb_2Te_3)_{1-x}$ structures with x ≤ 0.85 (Region 3).** a) Normalized temperature dependent Hall resistance ($R_{xy}$) taken at 200G for two samples of region 3, one grown with fast growth rates (red) and another one grown with slow growth rates (blue). The slow growth rate sample exhibited $T_C$ of about 75K. b) Superimposed plot of temperature dependent Hall resistance (dashed) and its derivative curve (solid) for the sample grown with a slow growth rate. Two $T_C$ components are seen: one at 20K and a second, broader one reaching as high as 80K. c) Field dependent Hall resistance ($R_{xy}$) for the slow growth rate sample, showing hysteresis up to 75K. d) Curie temperature $T_C$ plotted as a function of carrier density for samples grown at fast (red points) and slow (blue points) growth rates. No correlation between $T_C$ and carrier density is observed.

To understand the type of magnetic interactions that may be operating in these materials it would be useful to investigate the relationship between the carrier density and the $T_C$. In Figure 3d, we plot the $T_{C2}$ of samples with $x > 0.7$ (Regions 2 and 3) as a function of their carrier concentration. No apparent correlation between $T_C$ and carrier density is observed, suggesting the magnetic mechanism at play in these materials is not likely Ruderman–Kittel–Kasuya–Yosida (RKKY) interactions. These results are more consistent with magnetic behavior comparable to Van Vleck, or long-range magnetic interactions.[14, 28, 29, 30] Further investigations are needed to establish the types of magnetic interactions in these complex materials.

***Mn content of $(MnSb_2Te_4)_x(Sb_2Te_3)_{1-x}$ structures as a function of composition and growth rate: correlation with $T_C$ values.***

Excess Mn has been previously identified as a likely factor in determining the $T_C$ value of these MTIs. To explore this, we investigated the effects of growth rate on the elemental concentration of the $(MnSb_2Te_4)_x(Sb_2Te_3)_{1-x}$ samples using energy dispersive X-ray spectroscopy (EDS). The elemental contributions of Mn, Sb and Te were plotted as a function of composition (x) in the structure for a set of samples grown with fast and slow growth rates. The results are plotted in Figures 4 a and b. The solid black lines represent calculated values expected for stoichiometric $(MnSb_2Te_4)_x(Sb_2Te_3)_{1-x}$ layers, while the red data points represent the EDS data for a set of fast growth rate samples (from Ref 22), and the blue points, for the slow growth rate samples of this study. At low SL concentration, where $x \leq 0.7$, the data for fast growth rate samples largely follow the predicted curves for stoichiometric samples. However, when $x > 0.7$, the points begin to deviate significantly from the stoichiometric lines, showing increasing Mn content and a corresponding decrease in Sb content, consistent with a large excess of Mn occupying Sb sites. In the case of the slow growth rate samples, which all lie in this composition range of x ≥ 0.7, the excess Mn in the samples is even greater, and the Sb concentration is lower than for the fast growth rate samples. Furthermore, for the slow growth rate samples, some reduction in the Te concentration is also seen, suggesting that Mn is also substituting for Te atoms in crystal. Figure 4b shows a magnified view of the region of interest (the dashed square in Figure 4a) to illustrate this phenomenon more clearly. Thus, we conclude that the excess Mn in the crystal is greatly enhanced by the slow growth rates. It is important to note that it is within this composition range that we see evidence for the high $T_C$ in the samples. The relationship between $T_C$ and composition x can be clearly seen in Figure 4c. The data shows that the high $T_C$ for the fast growth rate samples and even higher $T_C$ for the slow growth rate samples occur when $0.7 \leq x \leq 0.85$, the same region where the excess Mn is present.

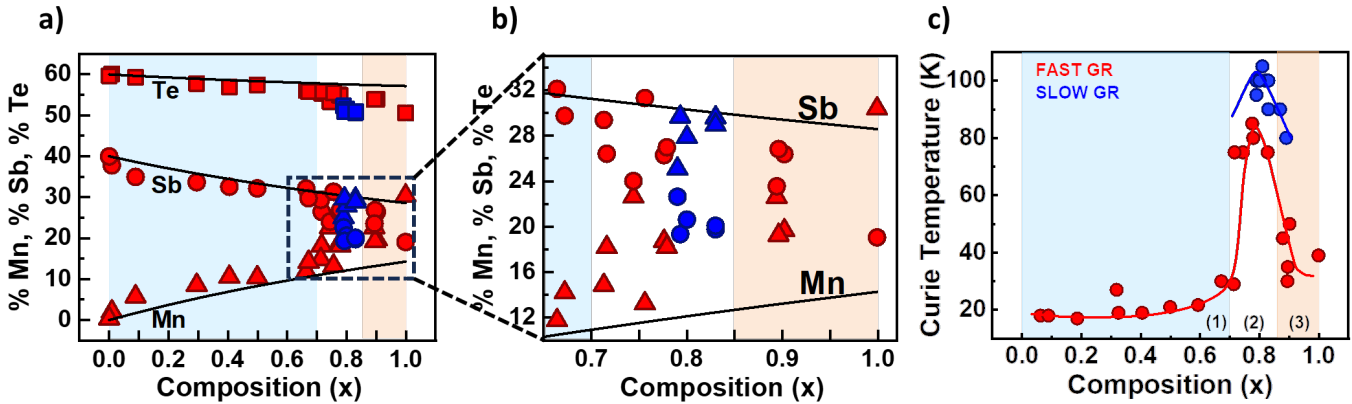

Figure 4: a) EDS characterization of a selection of samples grown at a fast growth rate (red) and slow growth rate (blue); solid lines represent composition for stoichiometric $(MnSb_2Te_4)_x(Sb_2Te_3)_{1-x}$. Site mixing between Mn and Sb is evident for fast growth rate samples with x above 0.7. Samples grown with the slower growth rate show a further increase in site mixing between Mn and Sb, as well as between Mn and Te atoms. b) A magnified view of the Sb and Mn EDS data for samples with x ≥ 0.7. c) Curie temperature as a function of composition x of $(MnSb_2Te_4)_x(Sb_2Te_3)_{1-x}$. Slow growth rate samples (blue points) consistently exhibit higher $T_C$ values for the same value of x.

The EDS results show that the reduced growth rate enhances the incorporation of Mn into the crystal structure. This can be understood if we consider that at the typical MBE growth temperatures, Sb desorption (and possibly some Te desorption), and its subsequent substitution by Mn, may be favored when growth takes place more slowly. Our magnetic investigations also showed that there are likely two different phenomena giving rise to the two different $T_C$ values. Thus, to explain these observations and the exceedingly high $T_C$ values that we observe, we propose that the excess Mn is incorporating in both the SL and the QL, allowing the formation of $Sb_{2-y}Mn_yTe_3$ alloy QLs, as well as Mn-rich $MnSb_2Te_4$ SLs. Mn-rich SLs have previously been shown[20,27] to yield FM materials, with $T_C$ as high as 40-50K, consistent with our observed $T_{C1}$. Other studies have also predicted high $T_C$ in $Sb_{2-y}TM_yTe_3$ alloys for TM = Cr, V, Mn, but the predictions have not been experimentally demonstrated for Mn due to a low Mn incorporation into $Sb_2Te_3$ under typical equilibrium crystal growth conditions.[24,25,26] We suggest that the non-equilibrium MBE growth conditions allow higher levels of Mn to incorporate into $Sb_{2-y}Mn_yTe_3$ alloys within our structure resulting in the observed higher $T_C$ values ($T_{C2}$).

Figure 5 illustrates the proposed structure for samples with $x \geq 0.7$, containing a mixture of QLs and SLs. For these compositions, growth by MBE requires very large Mn BEP ratios, and as shown by the EDS results, there is large excess of Mn in the crystal. In these samples, sufficient Mn is incorporated into the QL to form $Sb_{2-y}Mn_yTe_3$ alloys, which are predicted to be ferromagnetic with very high $T_C$ values. Since both Regions 2 and 3 have some QLs in the structure, there is evidence of a second $T_{C2}$ component in both regions. The Mn content y in the QLs is largest when the samples are grown at the slow growth rates, leading to the observed exceedingly high $T_{C2}$. The three frames seen in Figure 5 describe the temperature behavior of the FM properties of these structures. Below $T_{C1}$, SLs of composition $Mn_{1+y}Sb_{2-y}Te_4$ and QLs of composition $Mn_ySb_{2-y}Te_3$, where y represents the excess Mn in the structure, are magnetized and the structure exhibits strong FM behavior. Above the $T_{C1}$ but below $T_{C2}$, the spins are randomized in the SLs, and only the spins in the QL alloys are aligned, producing a weaker magnetization due to the reduced number of QLs in the structure, and leading to the high $T_{C2}$ that we observe in the $R_{xy}$ and M curves. Finally, above the $T_{C2}$, all spins are randomized. We suggest that high Mn content $Sb_{2-y}Mn_yTe_3$ QL is responsible for the high $T_{C2}$ values in our materials. Such high Mn content $Sb_{2-y}Mn_yTe_3$ alloys have not been previously realized experimentally and are formed here due to our far from equilibrium MBE growth conditions.

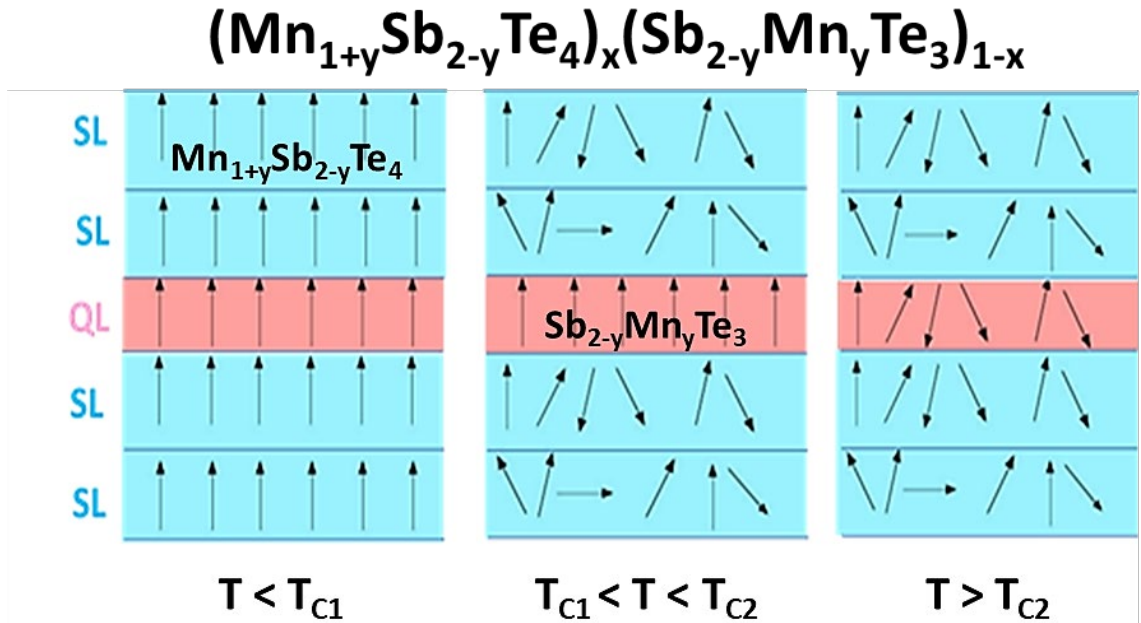

**Figure 5:** Schematic illustrating the origin of the two $T_C$ components in $(Mn_{1+y}Sb_{2-y}Te_4)_x(Mn_ySb_{2-y}Te_3)_{1-x}$ samples with $x \geq 0.7$ and the predicted temperature-dependent behavior of the Mn-spins for these samples.

**CONCLUSIONS**

We have shown that by reducing the growth rate during MBE growth, the highest $T_C$ value for $(MnSb_2Te_4)_x(Sb_2Te_3)_{1-x}$ structures reported to date, as high as 100K, was achieved. These values are much higher than those values reported by others,[20] and significantly higher than those recently reported by us,[27] for samples grown at faster growth rates. The values were measured by both Hall resistance and SQUID magnetization measurements yielding similar results. As in the fast growth rate samples previously reported, the results indicated the presence of two $T_C$ components. Structurally, no significant difference was apparent between samples grown at fast or slow growth rates. Only in the (0012) rocking curves did we see evidence of more disorder at the slow growth rates. Using EDS, we identified one important type of disorder that is present in these structures, which is enhanced by the slow growth rates. The elemental composition of these structures showed evidence of significant excess of Mn likely in the form of Mn:Sb and Mn:Te intermixing in both the SLs and the QLs. The enhanced Mn content is understood by recognizing that, using slower growth rates, there is a greater probability for Sb (and Te) desorption, which increases the likelihood of Mn incorporation in Sb sites and even in Te sites. The increased Mn content results in higher $T_C$ for samples grown with the same Mn BEP ratio. We propose that significant incorporation of Mn may be changing the $Sb_2Te_3$ QL into $Sb_{2-y}Mn_yTe_3$ alloys, which have been predicted to have very high $T_C$ values. Lastly, we illustrate the concept of two $T_C$ components in our structures, where $T_{C1}$ originates from the Mn-rich SLs and $T_{C2}$ originates from the Mn-rich $Sb_{2-y}Mn_yTe_3$ QLs. Preliminary results of the carrier density dependence on the $T_{C2}$ suggest that RKKY mechanism is not at play in these structures, and more long-range magnetic interactions are likely. These results propose a pathway to produce MTIs with even higher $T_C$ values by manipulating the growth conditions to form Mn-rich structures and bring us closer to demonstration practical applications based on these materials.

**ACKNOWLEDGMENTS**

This work was supported by NSF Grant no. DMR-2011738 (NSF MRSEC PAQM) and NSF Grant no. HRD-2112550 (Phase II CREST IDEALS). This work was also supported by Project DYN-TOP ANR-22-CE30-0026-01. The authors would like to acknowledge the Nanofabrication Facility of the CUNY Advanced Science Research Center (ASRC) for instrument use and scientific and technical assistance and the staff of the MPBT (physical properties—low temperature) platform of Sorbonne Universite for their support.

# Supplementary Information

# Structural and magnetic properties of molecular beam epitaxy $(MnSb_2Te_4)_x(Sb_2Te_3)_{1-x}$ topological materials with exceedingly high Curie temperature

Candice R. Forrester,[1,2,3] Christophe Testelin,[4] Kaushini Wickramasinghe,[1] Ido Levy,[5] Dominique Demaille,[4] David Hrabovski,[6] Xiaxin Ding,[7] Lia Krusin-Elbaum,[7,8] Gustavo E. Lopez,[2,3] Maria C. Tamargo[1,2]

[1]Department of Chemistry, The City College of New York, NY, NY 10031

[2]PhD Program in Chemistry, CUNY Graduate Center, NY, NY 10016

[3] Department of Chemistry, Lehman College, Bronx, NY 10468

[4]Sorbonne Université, CNRS, Institut des NanoScience de Paris, F-75005 Paris, France

[5]Department of Physics, New York University, NY, NY 10003

[6]Sorbonne Université, MPBT Platform, 4 Place Jussieu, 75252 Paris, France

[7]Department of Physics, The City College of New York, NY, NY 10031

[8] PhD Program in Physics, CUNY Graduate Center, NY, NY 10016

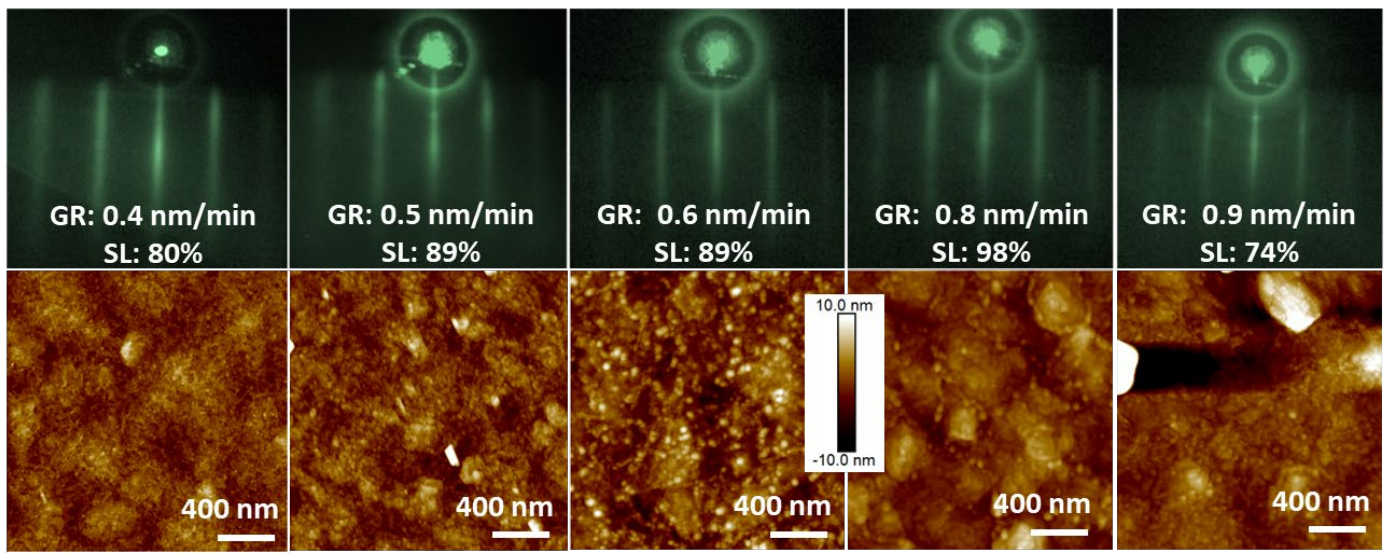

Figure S1: RHEED (top) and AFM images (bottom) of $(Sb_2Te_3)_{1-x}(MnSb_2Te_4)_x$ samples grown with increasing growth rates. Samples grown with a slower growth rate are on the left and samples grown with faster growth rates are on the right. No significant changes in the RHEED and AFM as the growth rate changed.

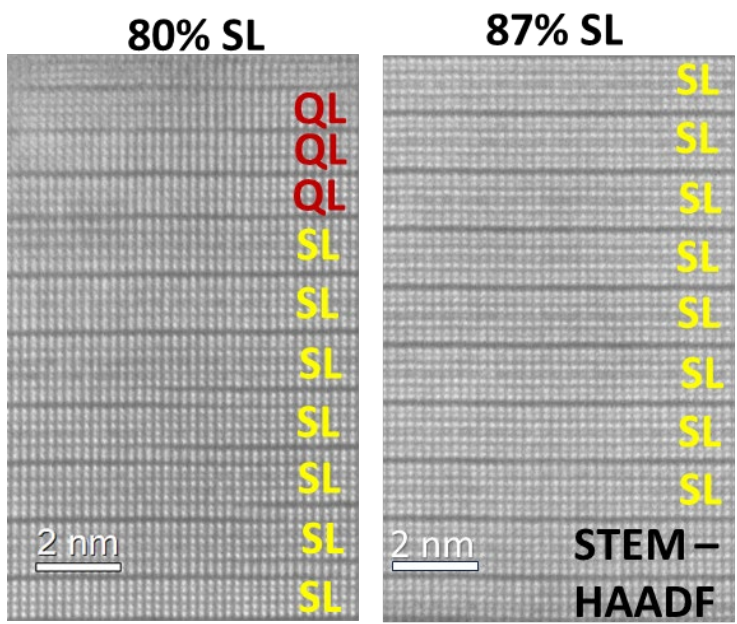

Figure S2: Cross-sectional STEM images of two samples with $x > 0.7$ grown with slow growth rates. Good crystalline quality is observed in $(Sb_2Te_3)_{1-x}(MnSb_2Te_4)_x$ samples

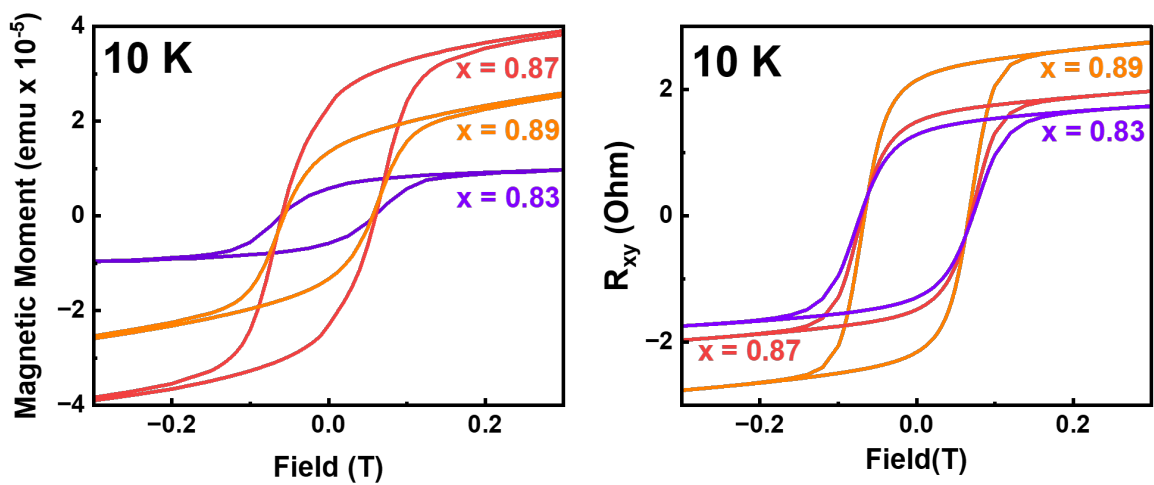

Figure S3: Field dependent magnetization (left) and Hall resistance (right) measurements taken at 10K for samples, where x ≥ 0.7. Hysteretic loops which indicate ferromagnetic behavior is consistently seen in both measurements. The sample where x = 0.83 is the same sample, exemplifying high $T_C$ in the manuscript (Figure 2e and 2f) and the sample, where x = 0.89 is also seen in Figure 3 in the manuscript.